# Intelligent Inverse Treatment Planning via Deep Reinforcement Learning, a Proof-of-Principle Study in High Dose-rate Brachytherapy for Cervical Cancer

Chenyang Shen, Yesenia Gonzalez, Peter Klages, Nan Qin, Hyunuk Jung, Liyuan Chen, Dan Nguyen, Steve B. Jiang, Xun Jia

Medical Artificial Intelligence and Automation Laboratory, Department of Radiation Oncology, University of Texas Southwestern Medical Center, Dallas, TX 75287, USA

E-mails: xun.jia@utsouthwestern.edu, chenyang.shen@utsouthwestern.edu

## Abstract

Inverse treatment planning in radiation therapy is formulated as solving optimization problems. The objective function and constraints consist of multiple terms designed for different clinical and practical considerations. Weighting factors of these terms are needed to define the optimization problem. While a treatment planning optimization engine can solve the optimization problem with given weights, adjusting the weights to yield a high-quality plan is typically performed by a human planner. Yet the weight-tuning task is labor intensive, time consuming, and it critically affects the final plan quality. An automatic weight-tuning approach is strongly desired. The procedure of weight adjustment to improve the plan quality is essentially a decision-making problem. Motivated by the tremendous success in deep learning for decision making with human-level intelligence, we propose a novel framework to adjust the weights in a human-like manner. This study uses inverse treatment planning in high-dose-rate brachytherapy (HDRBT) for cervical cancer as an example. We develop a weight-tuning policy network (WTPN) that observes dose volume histograms of a plan and outputs an action to adjust organ weighting factors, similar to the behaviors of a human planner. We train the WTPN via end-to-end deep reinforcement learning. Experience replay is performed with the epsilon greedy algorithm. After training is completed, we apply the trained WTPN to guide treatment planning of five testing patient cases. It is found that the trained WTPN successfully learns the treatment planning goals and is able to guide the weight tuning process. On average, the quality score of plans generated under the WTPN's guidance is improved by ~8.5% compared to the initial plan with arbitrarily set weights, and by 10.7% compared to the plans generated by human planners. To our knowledge, this is the first time that a tool is developed to adjust organ weights for the treatment planning optimization problem in a human-like fashion based on intelligence learnt from a training process. This is different from existing strategies based on pre-defined rules. The study demonstrates potential feasibility to develop intelligent treatment planning approaches via deep reinforcement learning.

## 1. INTRODUCTION

Inverse treatment planning is a critical component of radiation therapy (Oelfke and Bortfeld, 2001; Webb, 2003). It is typically formulated as an optimization problem, in which the objective function and constraints contain several terms designed for various clinical or practical considerations, such as dose volume criteria and plan deliverability. The optimization problem is solved mathematically to determine values of the set of variables defining a treatment plan, e.g. fluence map in external-beam radiation therapy (EBRT) and dwell time in high-dose-rate brachytherapy (HDRBT). These optimized values are further converted into control parameters of a treatment machine, namely a medical linear accelerator in EBRT and a remote afterloader in HDRBT, based on which the optimized treatment plan is delivered.

Mathematical formulation of the optimization problem in treatment planning typically contains a set of parameters to define different objectives. Examples of these parameters include, but are not limited to, positions and relative importance of different dose volume criteria. When adjusting these parameters, although the general formalism of the optimization problem remains unchanged, the resulting plan quality is affected. A modern treatment planning system can effectively solve the optimization problem with given parameters using a certain mathematical algorithm (Bazaraa *et al.*, 2006). Nonetheless, tuning these parameters for clinically satisfactory plan quality is typically beyond the capability of the algorithm. In a typical clinical setup, a human planner adjusts these parameters in a manual fashion. Not only does this prolong the treatment planning process, the final plan quality is affected by numerous factors, such as the experience of the planner and the available time on planning. Hence, there is a strong desire to develop automatic approaches to determine these parameters.

Over the years, extensive studies have been conducted to solve this parameter tuning problem. The most common approach is to add an additional iteration loop of parameter adjustment on top of the iteration used to solve the plan optimization problem with a fixed set of parameters. In a seminal study, Xing et. al. (Xing *et al.*, 1999) proposed to evaluate the plan quality in the outer loop and determine parameter adjustment using Powell's method towards optimizing the plan quality score. Similar approaches were taken by Lu et. al. using a recursive random search algorithm in intensity modulated radiation therapy (Lu *et al.*, 2007) and by Wu et. al using the genetic algorithm in 3D conformal therapy (Wu and Zhu, 2001). This two-loop approach was recently generalized by Wang et. al. to include guidance from prior plans designed for patients of similar anatomy. They also implemented the method in a treatment planning system to allow an automated planning process (Wang *et al.*, 2017). In the case with a large number of parameters in the optimization problem, e.g. one parameter per voxel, a heuristic approach was developed to adjust voxel-dependent parameters based on dose values of the intermediate solution (Yang and Xing, 2004; Wahl *et al.*, 2016) or based on the geometric information of the voxel (Yan and Yin, 2008). Other methods were also introduced to solve this problem. Yan et. al. employed a fuzzy inference technique to adjust the parameters (Yan *et al.*, 2003a; Yan *et al.*, 2003b). A statistical method was used by Lee et. al. (Lee *et al.*, 2013), which built the relationship between the parameters and the patient anatomy. Chan et. al. analyzed previously treated plans and developed a method to derive the parameters needed to recreate these plans. They further utilized statistical methods to establish a connection between patient anatomy and the optimal parameter set (Boutilier *et al.*, 2015; Chan *et al.*, 2014).

Parameter tuning in the plan optimization is essentially a decision making problem. Although it is difficult for a computer to automate this process, the task seems less of a problem for humans, as evidenced by the common clinical practice of manual parameter adjustment: a planner can adjust the parameters in a trial-and-error fashion based on human intuition. It is of interest and importance to model this remarkable intuition in an intelligence system, which can then be used to solve the parameter-tuning problem from a new angle. Recently, the tremendous success in deep-learning regime demonstrated that human-level intelligence can be spontaneously generated via deep-learning techniques. Pioneer work in this direction showed that a system built as such is able to perform certain tasks in a human-like fashion, or even better than humans. For instance, employing a deep Q-network approach, a system can be built to learn to play Atari games with a remarkable performance (Mnih *et al.*, 2015).

In fact, a human planner using a treatment planning system to design a plan is conceptually similar to a human playing computer games. Motivated by this similarity and the tremendous achievement in the deep learning area across many different problems (Mnih *et al.*, 2015; Silver *et al.*, 2016; Silver *et al.*, 2017; Chen *et al.*, 2018; LeCun *et al.*, 2015; Wang, 2016; Greenspan *et al.*, 2016; Zhen *et al.*, 2017; Nguyen *et al.*, 2017; Nguyen *et al.*, 2018; Shen *et al.*, 2018; Balagopal *et al.*, 2018; Iqbal *et al.*, 2017; Iqbal *et al.*, 2018; Ma *et al.*, 2018), we propose in this paper to develop an artificial intelligence system to accomplish the parameter-tuning task in an inverse treatment planning problem. Instead of tackling the problem in the EBRT context, we focus our initial study on an example problem of inverse planning in HDRBT with a tandem-and-ovoid (T/O) applicator for the purpose of proof of principles. This choice is made because of the relatively small problem size and therefore low computational burden. More specifically, based on an in-house optimization engine for HDRBT, we will build an intelligent system called Weight Tuning Policy Network to adjust organ weights in the optimization problem in a human-like fashion. The validity and generalization of this approach to the EBRT context will be discussed at the end of the paper.

## 2. METHODS AND MATERIALS

### 2.1 Optimization model for T/O HDRBT

Before presenting the system for organ weight tuning, we will first briefly define the optimization problem for T/O HDRBT. We considered an in-house developed optimization model (Liu *et al.*, 2017):

$$\min_{t} \sum_i \frac{\lambda_i}{2} \left\| M_{OAR}^i t \right\|_2^2 + \frac{1}{2} \|\nabla t\|_2^2 , \tag{1}$$
$$\text{s.t.} \, D^{CTV} = M_{CTV} t,$$
$$D^{CST} = M_{CST} t,$$

$$D^{CTV}(90\%) = D_p \, ,$$
$$D^{CST} \in [0.8\,D_p \, , \ 1.4\,D_p \,],$$
$$t_j \in [0, \ t_{max}], \ j = 1,2,\dots,n.$$

In this model, $M_{OAR}^i \in R^{m_i \times n}$ and $M_{CTV} \in R^{m_{CTV} \times n}$ are dose deposition matrices for the $i$-th organs at risk (OARs) and the clinical target volume (CTV). They characterize the dose to voxels in corresponding volumes of interest contributed from each dwell position at a unit dwell time. $m_i$, $m_{CTV}$, and $n$ are number of voxels in the OAR, that of the CTV, and the number of dwell positions, respectively. $t \in R^{n \times 1}$ is a vector of dwell time. The first term of the objective function minimizes the dose to OARs, and the regularization term $\|\nabla t\|_2^2$ enforces smoothness of the dwell time to ensure robustness of the resulting plan with respect to geometrical uncertainty of source positions. In addition, we impose the constraint to CTV, such that 90% of CTV volume should receive dose not lower than the prescription dose $D_p$. Moreover, according to the treatment planning guideline at our institution, a control structure (CST) is defined as two line segments that are parallel to the ovoid central axes and are on the outer surface of the ovoids. Dose in CST $D^{CST}$ should be within $[0.8D_p \, , \ 1.4D_p]$. The last constraint of the problem ensures that dwell time should be non-negative and less than a pre-defined maximum value. In this study, four OARs are considered, namely bladder, rectum, sigmoid and small bowel. $\lambda_i$s are the weights that control trade-offs among them. These organ weights determine the quality of the optimized plan, turning which is the interest of this paper.

For a given set of weights, we solve the optimization problem using the alternating direction method of multiplier (ADMM) (Boyd *et al.*, 2011). Here, we briefly present the algorithm and interested readers can refer to literature elsewhere (Liu *et al.*, 2017). The ADMM scheme allows us to tackle the problem via its augmented Lagrangian:

$$L(t,x,\Gamma) = \sum_i \frac{\lambda_i}{2} \left\| M_{OAR}^i t \right\|_2^2 + \frac{1}{2} \|\nabla t\|_2^2 + \frac{\beta}{2} \left\| \widehat{M}t - x \right\|_2^2 + \langle \Gamma, \widehat{M}t - x \rangle + \delta_1(x) + \delta_{box}(t), \tag{2}$$

where $\widehat{M} = \begin{pmatrix} M_{PTV} \\ M_{CST} \end{pmatrix}$ and $x = \begin{pmatrix} D^{PTV} \\ D^{CST} \end{pmatrix}$. $\Gamma$ indicates the Lagrangian multiplier and $\beta$ is the algorithm parameter to control the convergence. $\delta_1(x)$ and $\delta_{box}(t)$ are index functions that give 0 if constraints on $x$ and $t$ are satisfied, or $+\infty$ otherwise. The iterative process of the algorithm is summarized in Algorithm 1.

---

**Algorithm 1.** ADMM algorithm solving the problem in Eq. (1) with a given set of organ weights.

---

**Input:** $M_{OAR}^i$, $\widehat{M}$, $x^{(0)}, \Gamma^{(0)}$, $\lambda_i$, $\beta$ and tolerance $\sigma$

**Output:** $t^*$

**Procedure:**

1. Set $k = 0$;

2. Compute $t^{(k+\frac{1}{2})} = \left( \sum_i \lambda_i {M_{OAR}^i}^T M_{OAR}^i - \Delta + \beta \widehat{M}^T \widehat{M} \right)^{-1} \left( \beta \widehat{M}^T x^{(k)} - \widehat{M}^T \Gamma^{(k)} \right)$

3. Compute $t_j^{(k+1)} = \begin{cases} 0, & \text{if } t_i^{\left(k+\frac{1}{2}\right)} < 0 \\ t_{max}, & \text{if } t_i^{\left(k+\frac{1}{2}\right)} > t_{max} \ ; \\ t_i^{\left(k+\frac{1}{2}\right)}, & \text{otherwise} \end{cases}$

4.  Compute $x^{\left(k+\frac{1}{2}\right)} = \widehat{M}t^{(k+1)} + \frac{\Gamma^{(k)}}{\beta}, \begin{pmatrix} D^{CTV} \\ D^{CST} \end{pmatrix} = x^{\left(k+\frac{1}{2}\right)};$

5.  Compute $s = D^{CTV}(90\%), D_j^{CTV} = \begin{cases} D_p, & D_j^{CTV} \geq s \text{ and } D_j^{CTV} < D_p \\ D_j^{CTV}, & \text{otherwise} \end{cases}$ ,

    Compute $D_j^{CST} = \begin{cases} 0.8D_p, & \text{if } D_j^{CST} < 0.8D_p \\ 1.4D_p, & \text{if } D_j^{CST} > 1.4D_p \\ D_j^{CST}, & \text{otherwise} \end{cases}$ ,

    Compute $x^{(k+1)} = \begin{pmatrix} D^{CTV} \\ D^{CST} \end{pmatrix};$

6.  Compute $\Gamma^{(k+1)} = \Gamma^{(k)} + \beta(\widehat{M}t^{(k+1)} - x^{(k+1)})$

7.  If $\frac{\|t^{(k+1)}-t^{(k)}\|_2}{\|t^{(k+1)}\|_2} < \sigma$, set $t^* = t^{(k+1)}$;

    otherwise, set $k = k + 1$, go to Step 2.

### 2.2 Weight tuning methodology

We propose an automatic weight adjustment method for the aforementioned optimization engine by an artificial intelligence system that sequentially selects a weight and adjusts it. The system operates in a way analogous to the human-based treatment planning workflow: a planner repeatedly observes the plan obtained under a set of weights and makes a decision about weight adjustment, until a satisfactory plan quality is achieved (Fig. 1(a)). We aim at developing a Weight-Tuning Policy Network (WTPN) that serves the same purpose as a human planner in this workflow (Fig. 1(b)).

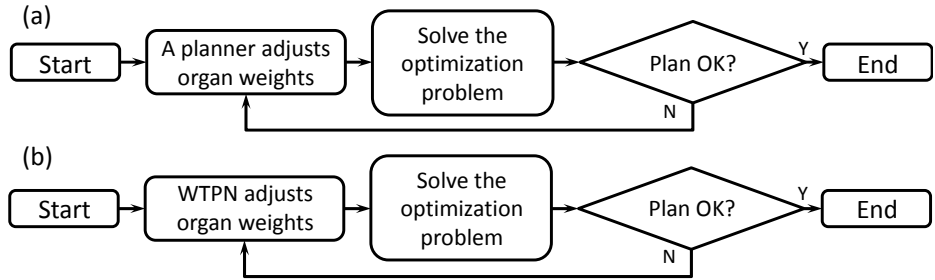

**Figure 1.** Illustration of weight tuning workflow (a) by a human planner and (b) by the WTPN.

More specifically, at the step $l$ of this weight tuning iteration, WTPN takes the dose-volume histograms (DVHs) in the plan as input and outputs a decision of weight adjustment: the organ weight to tune, and the direction and amplitude of the adjustment. Then, we update the weight and solve the optimization problem with the Algorithm 1. This process is repeated, until plan quality cannot be further improved.

To realize the proposed WTPN, we incorporate the Q-learning framework (Watkins and Dayan, 1992). This framework tries to build the optimal action-value function defined as:

$$Q^*(s,a) = \max_\pi [r^l + \gamma r^{l+1} + \gamma^2 r^{l+2} + \cdots | s^l = s, a^l = a, \pi]. \qquad (3)$$

$s$ is the current state, i.e. plan DVHs, and $s^l$ stands for the state at the $l$-th weight tuning step. $a$ is the action, i.e. which weight to adjust and how to adjust, and $a^l$ indicates the selected action. $r^l$ is the reward obtained at step $l$. In this study, the reward is calculated based on a pre-defined reward function related to clinical objectives. A positive reward is given, if the clinical objectives are better met by applying the action $a^l$ on the state $s^l$, and negative otherwise. $\gamma \in [0, 1]$ is a discount factor. $\pi = P(a|s)$ denotes the weight tuning policy: taking an action $a$ based on the observed state $s$. The goal of automatic weight tuning is to build the $Q^*$ function. Once this is achieved, the policy is determined as choosing the action that maximizes the $Q^*$ function value under the observed state $s$, i.e. $a = \arg \max_{a'} Q^*(s, a')$.

The form of the $Q^*$ function is generally unknown. In this paper, we propose to parametrize $Q^*$ via a deep convolutional neural network (CNN), denoted as $Q(s, a; W)$. $W = \{W_1, W_2, \dots, W_N\}$ indicates the network parameters. The network consists of $N$ independent subnetworks (see Fig. 2(a)) one for an OAR weight. The subnetworks share the same structure as displayed in Fig. 2(b). We defined five possible tuning actions for each weight: increase or decrease the weight by 50%, increase or decrease the weight by 10%, and keep the weight unchanged. The values 50% and 10% are arbitrary chosen, as we expect they would not critically affect the capability of weight tuning but only the speed to reach convergence. Each subnetwork has five outputs. The network takes observed state $s$, i.e. DVHs as input, and outputs values of the $Q$ function at each output node, corresponding to an action. The parameters $W_i$ of each network will be determined via the reinforcement learning strategy presented in the next section.

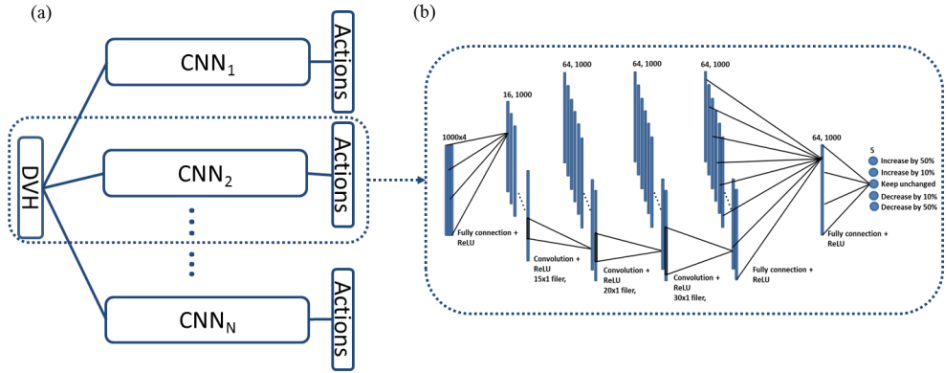

**Figure 2.** Network structure of the WTPN. (a) gives the overall structure of WTPN. The complete network consists of **$N$** subnetworks with identical structures. Each subnetwork corresponds to one OAR. The input is DVHs of a treatment plan. (b) Detailed structure of the subnetwork. Numbers and sizes of different layers are specified at the top of the layer and connections between layers are presented at the bottom. Output value of each network node is the corresponding **$Q$** function value of defined action.

### 2.3 Deep reinforcement learning

#### 2.3.1 General idea of network training

The training process is based on Bellman equation (Bellman and Karush, 1964), which is a general property satisfied by the optimal action-value function $Q^*(s, a)$:

$$Q^*(s, a) = r + \gamma \max_{a'} Q^*(s', a'), \tag{4}$$

where $r$ is the reward after applying action $a$ to the current state $s$ and $s'$ is the state after taking the action $a$. Using a CNN $Q(s, a; W)$ as an approximation of the $Q$ function, we define a quadratic loss function with respect to the network parameter $W$:

$$H(W) = \left[ r + \gamma \max_{a'} Q(s', a'; W) - Q(s, a; W) \right]^2.$$ (5)

Our goal is to determine $W$ through a reinforcement learning strategy to minimize this loss function, which hence ensures Eq. (4) and therefore $Q(s, a; W)$ will approach the optimal action-value function $Q^*(s, a; W)$. More specifically, let $\hat{W}$ denote a set of CNN parameters, $W$ is updated by minimizing the following loss function with $\hat{W}$ fixed:

$$L(W) = \left[ r + \gamma \max_{a'} Q(s', a'; \hat{W}) - Q(s, a; W) \right]^2.$$ (6)

The learning process consists of a sequence of stages. At each stage, $W$ is calculated to minimize $L(W)$ with the stochastic gradient descent method and fixed $\hat{W}$. The gradient of the loss function $L(W)$ can be simply derived as

$$\frac{\partial L(W)}{\partial W} = \left[ Q(s, a; W) - r + \gamma \max_{a} Q(s', a'; \hat{W}) \right] \frac{\partial Q(s, a; W)}{\partial W},$$ (7)

where the last term $\partial Q(s, a; W) / \partial W$ can be computed via the standard back-propagation strategy (LeCun *et al.*, 1998). With the gradient of loss function ready, $W$ at each step can be updated by a gradient descent form:

$$W = W - \delta \frac{\partial L}{\partial W},$$ (8)

where $\delta$ is the step size. We use stochastic gradient descent that computes the gradient and updates $W$ with a subset of the training data randomly selected from the training data set. After finishing each stage of training, $\hat{W}$ is updated by letting $\hat{W} = W$ and then fixed for the next stage of training. Eventually, $\hat{W}$ and $W$ are expected to converge at the end of the learning process.

### 2.3.2. Reward function

One important issue is to quantitatively evaluate the plan quality. In general, this is still an open problem and different evaluation metrics can be proposed depending on the clinical objectives. In our case, since the plan is always normalized to $D^{CTV}(90\%) = D_p$ in the optimization algorithm (Algorithm 1), we consider OAR sparing to assess the plan quality, as quantified by $D_{2cc}$ in the HDRBT context (Viswanathan *et al.*, 2012). For simplicity, we measure the plan quality as $\psi = \sum_i \omega_i D_{2cc}^i$ where $\omega_i$ are the preference factors indicating the radiation sensitivity of the $i$-th OAR. the lower $\psi$ is, the better plan quality is. In principal, a larger $\omega_i$ should be assigned to a more radiation sensitive OAR. We then formulate the following reward function regarding the change of from state $s$ to $s'$:

$$\Phi(s, s') = \psi(s) - \psi(s') = \sum_i \omega_i(D_{2cc}^i(s) - D_{2cc}^i(s')).$$ (9)

$s$ indicates the state (DVHs) prior to weight adjustment, while $s'$ is that after. The reward $\Phi(s, s')$ explicitly measures the difference in plan quality between the two states. $\Phi(s, s')$ is positive if plan quality is improved, and negative otherwise.

### *2.3.3 Training strategy*

The training process is performed in a number of $N_{episode}$ episodes. Each episode contains a sequence of $N_{train}$ steps indexed by $l$. At each step, we select an action to adjust an OAR's weight using the $\epsilon$-greedy algorithm. Specifically, with a probability of $\epsilon$, we randomly select one of the OARs and one action to adjust its weight. Otherwise, the action $a$ that attains the highest output value of the network $Q(s, a; W)$ is selected, i.e. $a^l = \arg\max_a Q(s^l, a; W)$. After that, we apply the selected action to the corresponding OAR's weight and solve the plan optimization problem of Eq. (1) using the Algorithm 1, yielding a new plan with DVHs denoted as $s^{l+1}$. $s^l$ and $s^{l+1}$ are then fed into the reward function $\Phi$ defined in Eq. (9) to calculate $r^l$.

At this point, we collect $\{s^l, a^l, r^l, s^{l+1}\}$ into the pool of training data set for the network $Q$. $W$ is then updated by the experience replay strategy to minimize the loss function in Eq. (6) using a number of $N_{batch}$ training samples randomly selected from the training data pool via Eq. (8). The main purpose of this experience replay strategy is to overcome the strong correlation among the sequentially generated training samples described in the last paragraph (Mnih *et al.*, 2015). Once the maximum number of training step $N_{train}$ is reached, we move to the next patient and apply the above training process again. Within this process, $\widehat{W}$ is updated by letting $\widehat{W} = W$ at every $N_{update}$ steps. The complete structure of the training framework is outlined in Algorithm 2.

---

**Algorithm 2.** Overall algorithm to train the WTPN.

---

Initialize network coefficients $W$;
**for** episode $= 1, 2, \ldots, N_{episode}$
    **for** $k = 1, 2, \ldots, N_{patient}$ **do**
        Initialize $\lambda_1, \lambda_2, \ldots, \lambda_N$;
        Run Algorithm 1 with $\{\lambda_1, \lambda_2, \ldots, \lambda_N\}$ for $s^1$;
        **for** $l = 1, 2, \ldots, N_{train}$ **do**
            Select an action $a^l$:
                **Case 1:** with probability $\epsilon$, select $a^l$ randomly;
                **Case 2:** otherwise $a^l = \arg\max_a Q(s^l, a; W)$;
            Based on selected $a^l$, adjust corresponding organ's weight;
            Run Algorithm 1 updated weights for $s^{l+1}$;
            Compute reward $r^l = \Phi(s^l, s^{l+1})$;
            Store reward $\{s^l, a^l, r^l, s^{l+1}\}$ in training data pool;
            Train $W$:
                Randomly select $N_{batch}$ training data from training data pool;
                Compute gradient using Eq. (7);
                Update $W$ using Eq. (8);
            Set $\widehat{W} = W$ every $N_{update}$ steps;
        **end for**
    **end for**
**end for**

---

The WTPN framework is implemented using Python with TensorFlow (Abadi *et al.*, 2016) on a desktop workstation equipped with eight Intel Xeon 3.5 GHz CPU processors, 32 GB memory and two Nvidia Quadro M4000 GPU cards. We use five patient cases in training. Note that the data to train the WTPN are in fact $\{s^l, a^l, r^l, s^{l+1}\}$ generated in the process outlined above. With five patient cases, we are able to generate enough training data. The initial weights for all OARs are set to unity. Other major hyperparameters to configure our system are summarized in Table 1.

**Table 1.** Hyperparameters to train the weight tuning system.

| Hyperparameter | Value | Description |
|---|---|---|
| $\sigma$ | $5 \times 10^{-4}$ | Stopping criteria in Algorithm 1 |
| $\beta$ | 5 | Penalty parameter in Algorithm 1 |
| $n$ | 4 | Number of weights (OARs) to be tuned |
| $\gamma$ | 0.5 | Discount factor |
| $\epsilon$ | $0.99 \sim 0.1$ | Probability of $\epsilon$-greedy approach |
| $N_{episode}$ | 300 | Number of training episodes |
| $N_{train}$ | 30 | Number of training steps in each episode |
| $N_{update}$ | 10 | Number of steps to update $\widehat{W}_i = W_i$ |
| $\delta$ | $1 \times 10^{-4}$ | Learning rate (step size of gradient descent for $W_i$) |

## 2.4 Validation studies

The WTPN is developed to adjust organ weights to gain a high reward $\Phi$, which will improve the plan quality, as quantified by reduction of $\psi = \sum_i \omega_i D_{2cc}^i$. To validate the WTPN, we use the trained WTPN to adjust OAR weights in those five cases used in training and five additional independent testing cases. Without loss of generality, we set $\omega_{bladder} = 0.2$ while $\omega_{rectum} = \omega_{sigmoid} = \omega_{smallbowel} = 1$ in the reward function $\Phi$, as bladder is more radiation resistant compared to the other OARs. In each case, we perform the weighting adjustment process using the trained WTPN as shown in Fig. 1(b). Evolution of the plan quality in this process is studied.

In addition, we also train and test another WTPN using the preference factors $\omega_i = 1$, for $i = 1, \dots, 4$. The plan quality metric in this case is denoted as $\hat{\psi}$. The purpose of this study is to demonstrate the capability of adapting the developed method to different plan quality metrics. Clinically, different plan quality metrics can be interpreted as different preference of organ trade-offs, probably due to different physicians. It is important to study the adaptability of the proposed scheme to ensure clinical utility. Additionally, we compare the performances of the two WPTNs trained with $\psi$ and $\hat{\psi}$ functions.

## 3. RESULTS

### 3.1 Training process

The recorded reward and Q-values along training epochs are displayed in in Fig. 3. Note that reward reflects the plan score obtained via automatic weight tuning using WTPN, while the Q-value indicates output of WPTN approximating future rewards to be gained via weight adjustment. It can be observed in Fig. 3 that the reward and Q-value both show increasing trends, indicating that the WTPN gradually learns a policy of weight tuning that can improve the plan quality.

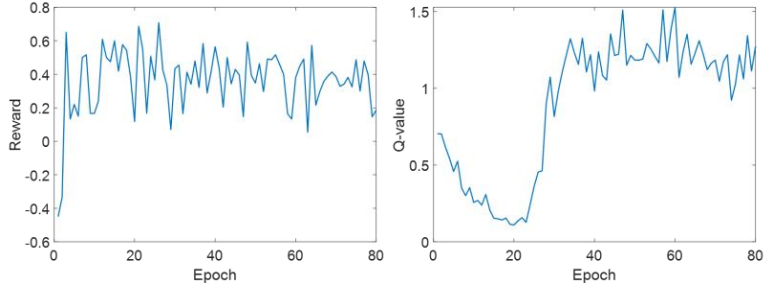

**Figure 3.** Reward (left) and Q-values (right) obtained along training epochs.

### 3.2 Weight tuning process

In Fig. 4, we present how the trained WTPN performs weight adjustment in an example case 3 that is used in training. Fig. 4(a) shows evolution of the weights. Corresponding $D_{2cc}$ values of different OARs are displayed in Fig. 4(b), which provide insights of how the proposed WTPN performs weight adjustment. In the initial eight steps, WTPN first increases the rectum weight, resulting in a successful reduction of $D_{2cc}^{rectum}$ at the expense of increasing $D_{2cc}^{sigmoid}$ and $D_{2cc}^{small\ bowel}$. $D_{2cc}^{bladder}$ is first reduced and later increased. The $\psi$ function value is greatly reduced. From step 8 to 12, the bladder weight is reduced, allowing reduction of other organ doses and slightly reduction of $\psi$. Starting from step 12, WPTN decides to increase the sigmoid weight probably due to the observed large $D_{2cc}^{smallbowel}$. Overall, the $\psi$ function value shows an overall decreasing trend, which indicates that the plan quality is significantly improved under the guidance of WTPN. The final $\psi$ value is lower than that of the clinical plan that was used in our clinic to treat this patient. In addition, we plot the DVHs at tuning steps 0 (initial), 5, and 25 in Fig. 4(d), while DVHs plotted with absolute volume around 2cc are shown in 4(e).

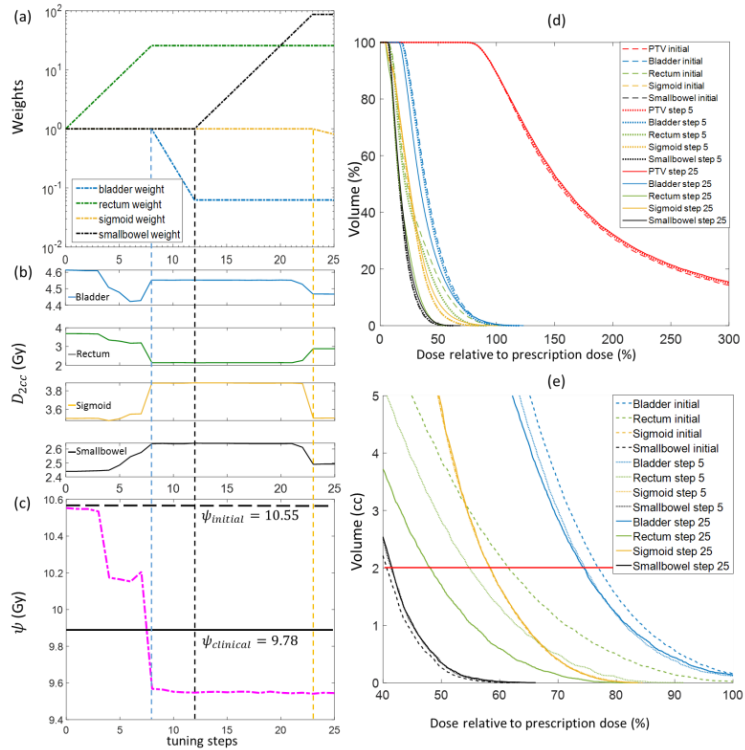

**Figure 4.** (a) Evolution of organ weights for training case 3; (b) Corresponding $\boldsymbol{D_{2cc}}$ of different OARs; (c) $\boldsymbol{\psi}$ function values; (d) DVHs of plans at weight tuning steps 0 (initial weights), 5 and 25; (e) DVHs plotted with absolute volume. Horizontal line shows 2cc volume.

Similarly, we show in Fig. 5 the weight-tuning process for the testing case 3 that is not included in the training of the WTPN. For this case, WPTN decides to first increase rectum weights, causing reduced $D_{2cc}$s for bladder, rectum, and sigmoid. Starting from step 15, WTPN increases small bowel weight. Dose to small bowel is successfully reduced without affecting too much on dose to rectum and sigmoid. $D_{2cc}^{bladder}$ increases a little, which is reasonable as it is our assumption that bladder is more radiation resistant (with a lower preference factor of 0.2). In general, the $\psi$ function value, as well as dose to OARs for this testing case have been successfully reduced in this process.

### 3.3 All training and testing cases

We report the performance of WTPN on the five training and five testing cases in

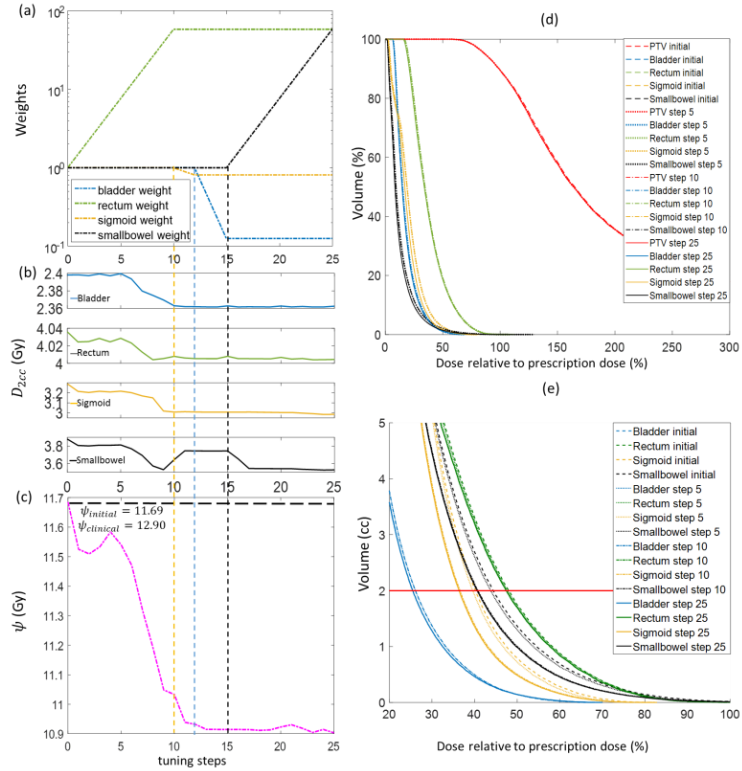

**Figure 5.** (a) Evolution of organ weights for testing case 3; (b) Corresponding $D_{2cc}$ of different OARs; (c) $\psi$ function values; (d) DVHs of plans at weight tuning steps 0 (initial weights), 5 and 25; (e) DVHs plotted with absolute volume. Horizontal line shows 2cc volume.

Table 2. Consistent improvements are observed for all the cases compared to those plans generated with initial weights. The plans after weight tuning are also better than those manually generated by the planners in our clinic.

**Table 2.** Weight tuning results for training cases. Numbers in bold face are the smallest values in each case.

| Cases | $\psi_{initial}$ (Gy) | $\psi_{tuned}$ (Gy) | $\psi_{clinical}$ (Gy) |
|---|---|---|---|
| Training patient 1 | 6.53 | **6.17** | 6.62 |
| Training patient 2 | 8.37 | **7.31** | 8.28 |
| Training patient 3 | 10.55 | **9.35** | 9.78 |
| Training patient 4 | 10.72 | **10.54** | 10.79 |
| Training patient 5 | 6.18 | **5.82** | 6.19 |
| Testing patient 1 | 6.81 | **6.48** | 6.61 |
| Testing patient 2 | 5.95 | **5.07** | 6.13 |
| Testing patient 3 | 11.69 | **10.90** | 12.90 |
| Testing patient 4 | 9.74 | **8.94** | 10.02 |
| Testing patient 5 | 10.18 | **9.19** | 9.78 |

For all the training cases, on average the $\psi$ function values after automatic weight tuning are reduced by 0.63 Gy (~7.5%) compared to the initial plans, and 0.50 Gy (~6%) compared to clinical plans. In the testing cases, average $\psi$ values under WPTN guidance are 0.76 Gy (~8.5%) and 0.97 Gy (~10.7%) lower than those of the initial plans and of the clinical plans, respectively. These numbers clearly demonstrate the effectiveness of the developed WPTN.

To get a better understanding on the plan quality, we use the testing patient 5 as an example and show its DVH curves of the initial plan, clinical plan and automatically tuned plan in Fig. 6. It is clear that doses to rectum, sigmoid and small bowel are effectively reduced by the WPTN. Among them, the DVH curves for sigmoid and small bowel obviously outperform the clinical plan. The dose to bladder is higher than that under the initial organ weight setup. Due to the assumption that bladder is more radiation resistant compared to the other OARs ($\omega_{bladder} = 0.2$), WPTN decides to sacrifice bladder to reduce $\psi$ and hence increase plan quality.

The advantage of WPTN can be also observed directly on isodose lines. Using the testing patient 2 as an example, the OARs are spared successfully, especially in the highlighted areas indicated by pink circles in Fig. 7. More specifically, it is shown in coronal view (Fig. 7(a)) that the dosages to small bowel, sigmoid and rectum using WPTN are apparently the lowest among the three plans. Similarly, in Fig. 7(b) sigmoid and small bowel receive lower dose in the weight-adjusted plan than the other two plans. Note that all these cases have the same CTV coverage of $D^{CTV}(90\%) = D_p$ because of the constraint in the optimization problem.

### 3.4 Impact of preference factors in reward function

Table 3 reports weight tuning results using $\hat{\psi}$ in the reward function, in which the preference factors for all the OARs are set to unity. After training the WPTN with the new reward function, WPTN is again able to successfully adjust OAR weights of the objective function, so that the values $\hat{\psi}$ are reduced through the planning process. The resulting $\hat{\psi}$ at the end are lower than those in the clinical plan, indicating better plan quality.

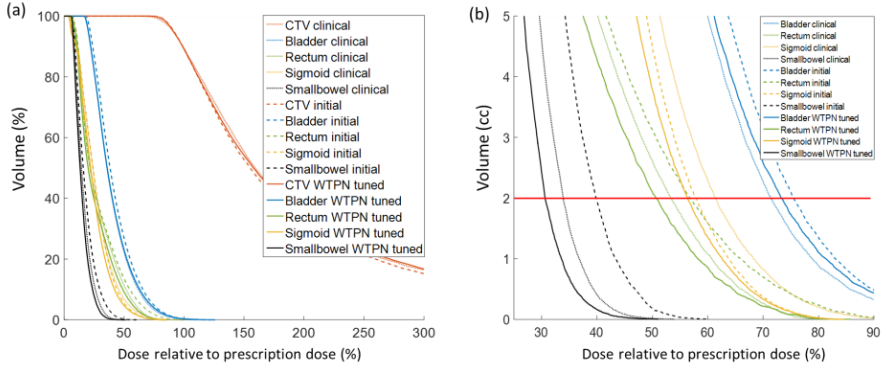

**Figure 6.** DVH comparison curves for testing patient case 5.

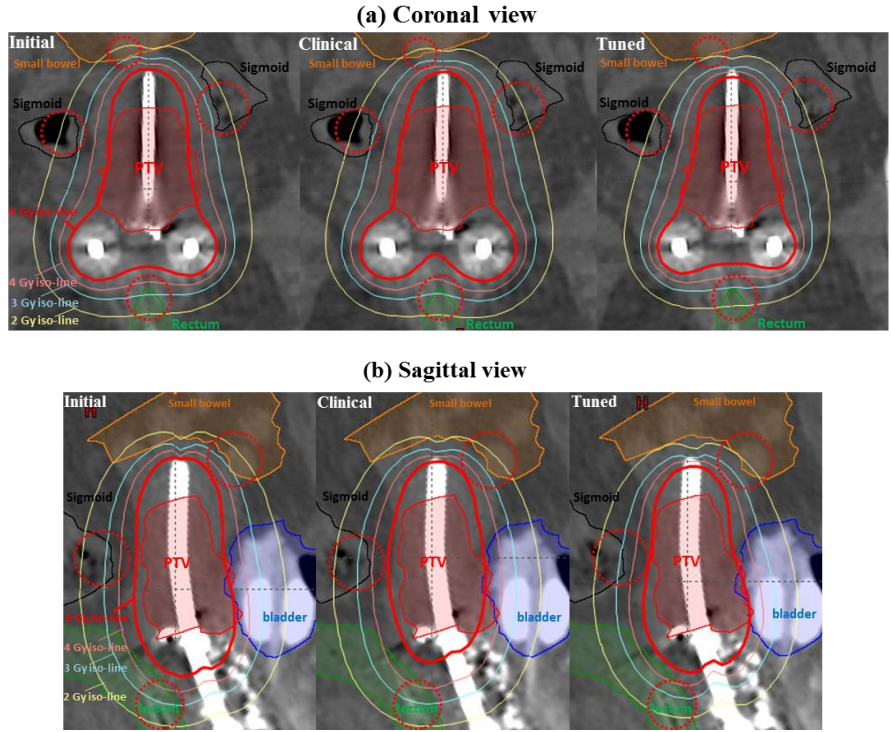

**Figure 7.** Dose map comparison for patient case 2.

Table 4 compares plan results generated by WTPN with two different reward functions using $\psi$ and $\hat{\psi}$. Note that the difference between the two setups is that bladder is considered to be more important in $\hat{\psi}$ ($\omega_{bladder} = 1$). In response to the increased preference factor for bladder, the resulting plan has a lower bladder $D_{2cc}$. At the same time, other OARs are affected to different degrees. $D_{2cc}$ of them are mostly increased when $\hat{\psi}$ is used because of the consideration of bladder sparing.

Table 3. **Weight tuning results for testing cases.** Numbers in bold face are the smallest values in each case.

| Cases | $\hat{\psi}_{initial}$ (Gy) | $\hat{\psi}_{tuned}$ (Gy) | $\hat{\psi}_{clinical}$ (Gy) |
|---|---|---|---|
| Testing patient 1 | 10.03 | **9.40** | 9.75 |
| Testing patient 2 | 6.85 | **6.45** | 7.17 |
| Testing patient 3 | 13.60 | **13.31** | 15.47 |
| Testing patient 4 | 13.39 | **12.79** | 13.94 |
| Testing patient 5 | 13.80 | **12.75** | 13.21 |

**Table 4.** Effect of different reward functions on testing cases.

| Cases | Reward | $D_{2cc}^{bladder}$ (Gy) | $D_{2cc}^{rectum}$ (Gy) | $D_{2cc}^{sigmoid}$ (Gy) | $D_{2cc}^{smallbowel}$ (Gy) |
|---|---|---|---|---|---|
| Testing patient 1 | $\psi$ | 3.89 | 2.55 | 2.09 | 1.06 |
| | $\hat{\psi}$ | 3.76 | 2.56 | 2.08 | 1.00 |
| Testing patient 2 | $\psi$ | 1.18 | 1.35 | 2.89 | 0.59 |
| | $\hat{\psi}$ | 0.97 | 1.70 | 3.12 | 0.66 |
| Testing patient 3 | $\psi$ | 2.51 | 3.96 | 2.95 | 3.49 |
| | $\hat{\psi}$ | 2.38 | 4.01 | 3.18 | 3.74 |
| Testing patient 4 | $\psi$ | 4.56 | 3.29 | 2.13 | 2.60 |
| | $\hat{\psi}$ | 4.45 | 3.41 | 2.19 | 2.74 |
| Testing patient 5 | $\psi$ | 4.47 | 3.06 | 3.37 | 1.87 |
| | $\hat{\psi}$ | 4.41 | 3.09 | 3.39 | 1.86 |

## 4. DISCUSSIONS

As mentioned in the introduction section, a representative approach in existing efforts to adjust weighting factors in the treatment planning optimization problem is to add a second loop on top of the iteration of solving the plan optimization problem. In each step, the weights are adjusted based on certain mathematical rules aiming at improving the plan quality, as quantified by a certain metric (Xing *et al.*, 1999; Lu *et al.*, 2007; Wu and Zhu, 2001; Wang *et al.*, 2017). Compare to these approaches, our method attained a similar structure, in the sense that the OAR weights are adjusted in an iterative fashion in the outer loop. Nonetheless, a notable difference is that, in contrast to previous approaches adjusting weights by a certain rigorous or heuristic mathematical algorithms, our system is designed and trained to develop a policy that can intelligently tune the weights, akin to the behavior of a human planner. The reward function involving the plan quality metric is only used in the training stage to guide the system to generate the intelligence. When WTPN is trained, the goal of treatment planning, i.e. to improve plan quality metric, is understood and memorized by the system. The subsequent application of the WTPN to a new case does not explicitly operate in a way aiming at mathematically improving the plan quality metric. Instead, WTPN behaves with the learnt intention to improve the plan, as having been clearly demonstrated in the testing studies.

The WTPN system is developed under the motivation to represent the clinical workflow, in which a planner repeatedly tunes the organ weights based on human intuition to improve the clinical objective. The WTPN, once is trained, assumes the planner's role in this workflow (Fig. 1). Yet, one apparent issue is that the developed system now becomes a black box and it is difficult to interpret the reasons for weight adjustments. Therefore, it is difficult to justify the rigor of the approach. All that can be shown is that the trained WTPN appears to be able to work in a human-like manner. In fact, it is a central topic in the deep learning area to decipher the underlying intelligence in a trained system (Zhang *et al.*, 2016; Zhang *et al.*, 2018; Che *et al.*, 2016; Sturm *et al.*, 2016). It will be our ongoing work to pursue this direction, which is essential for a better understanding of the developed system, for further improving its performance, and for its safe clinical implementation.

This study selects a problem of inverse optimization in T/O HDRBT instead of a more commonly studied problem of EBRT. This is for the consideration of using a relatively simple problem with a small problem size to reduce the computational burden. Despite this limitation, it is conceivable that the proposed approach is generalizable to the optimization problem in EBRT. In fact, the method described in Section 2 has a rather generic structure that takes an intermediate plan as input and outputs the way to change parameters in the optimization problem. It does not depend on the specific optimization problem of interest. Nevertheless, we admit that generalization of the proposed method to the EBRT regime will encounter certain difficulties. Not only will the optimization problem itself be substantially larger in size, which will inevitably prolongs computation time each time solving the optimization problem, the number of parameters to tune will also be much larger in an EBRT problem. The latter issue will lead to a much larger WTPN to train, which will hence cause a larger computational burden to train the network. We also envision that, in the EBRT regime, justifying a plan quality is a much complex problem than in that of HDRBT. This will yield the challenge of properly defining the reward function, i.e. a counterpart of Eq. (8) in EBRT. It will be our future study to extend the proposed approach to EBRT, as well as to overcome the aforementioned challenges.

One advantage of the proposed method is that it naturally works on top of any existing optimization systems. Similar to the study by Wang et. al. (Wang *et al.*, 2017), the developed system can be partnered with an existing treatment planning system (TPS). The only requirement is that the TPS has an interface to allow querying a treatment plan and inputting updated weights to launch an optimization, which is already feasible in many modern TPSs, for instance Varian Eclipse API (Varian Medical Systems, Palo Alto, CA). In addition, one notable fact in the proposed approach is it takes a plan that is generated by an optimization engine as input. This could be the plan after all required processing steps by the TPS, for instance after leaf sequencing operations in an EBRT problem. This fact is has practical benefits, as it can address the subtle quality difference in a plan caused by the leaf sequencing operations. In contrast, if we were to directly add a layer of weight optimization to the plan optimization by solving the problem from a mathematically rigorous way, it would be difficult to derive operations to account for this difference. Heuristic approach would likely have to be used.

Another, but more straightforward way to determine the weights in the deep learning context is to use a large number of optimized cases to build a connection between patient anatomy and the optimal weights. This is in fact the mainstream of those existing applying deep learning techniques to solve a spectrum of problems in medicine. Yet one drawback is the requirement on the number of training cases. The number necessary to build a reliable connection is typically very large, posing a practical challenge. In contrast, our study is motivated by mimicking human behaviors. In fact, the key behind the reinforcement learning process is to let the WTPN to try different parameter tuning strategies in the $\epsilon$-greedy algorithm, differentiate between proper and improper ways of adjustment, and memorize those proper ones. This is similar to teaching a human planner to learn how to develop a high-quality plan. As demonstrated in our studies, one apparent advantage is that, with a relatively low number of patient cases, successful training can be accomplished. It is also noted that the actual data to train WTPN are not the patient cases, but the state-action pair $\{s^l, a^l, r^l, s^{l+1}\}$ generated in the reinforcement learning process. If we count the state-action pair data, the number of training data is in fact large.

The current study is for the purpose of proof of principle and has the following limitations. First, the reward function may not be clinically realistic. The choice of Eq. (8) was a simple one that reflects physician's idea to a certain extent in HDRBT. By no means it should be interpreted as the one used in a real clinical situation. However, we also point out that the reward function in our system can be changed to any quantities based on clinical or practical considerations. In essence, the system is developed to mimic the human planner's behavior in the clinical treatment planning workflow. Hence, the reward function here is akin to a metric to quantify the physician's judgement of a plan. In the past, there have been several studies aiming at developing such a metric (Moore *et al.*, 2012; Zhu *et al.*, 2011). In principle, these metrics can be used in our system. In addition, recent advancements in imitation learning and inverse deep reinforcement learning (Wulfmeier *et al.*, 2015) allow learning the reward function based on human behavior. In the treatment planning context, it may be possible to learn the physician's preference as represented by the reward function. It is our ongoing work to perform studies as such.

Another limitation is that WTPN only takes DVH as input, which hence neglects other aspects of a plan. For instance, in an EBRT problem, DVH cannot capture position-specific information such as locations of hot/cold spots, which a physician often pays attention to. Again, at this early stage of developing an human-like intelligence system for weight tuning, we made the decision to start with a relatively simple setup to illustrate our idea. Further extending the system to include more realistic and clinically important features will be down the road.

## 5. CONCLUSION

In this paper, we have proposed a deep reinforcement learning-based weight tuning network WTPN for inverse planning of radiotherapy. We chose the relatively simple context of T/O HDRBT to demonstrate the principles. The WTPN was constructed to decide organ weight adjustments based on observed DVHs, similar to the behaviors of a human planner. The WTPN was trained via an end-to-end reinforcement learning procedure. When applying the trained WTPN, the resulting plans outperformed those plans optimized with initial weights significantly. Compared to the clinically accepted plans made by human planers, WTPN generated better plans with same CTV coverage in all the testing cases. To our knowledge, this was the first time that an intelligent tool is developed to adjust organ weights in a treatment planning optimization problem in a human-like fashion based on intelligence learnt from a training process, which is fundamentally different from existing strategies based on pre-defined rules. Our study demonstrated potential feasibility to develop intelligent treatment planning approaches via deep reinforcement learning.

**Reference**

Abadi M, Barham P, Chen J, Chen Z, Davis A, Dean J, Devin M, Ghemawat S, Irving G and Isard M *OSDI,2016),* vol. Series 16*)* pp 265-83

Balagopal A, Kazemifar S, Nguyen D, Lin M-H, Hannan R, Owrangi A and Jiang S 2018 Fully Automated Organ Segmentation in Male Pelvic CT Images *arXiv preprint arXiv:1805.12526*

Bazaraa M S, Sherali H D and Shetty C M 2006 *Nonlinear Programming: Theory and Algorithms*: Hoboken: A John Wiley & Sons)

Bellman R and Karush R 1964 Dynamic programming: a bibliography of theory and application. RAND CORP SANTA MONICA CA)

Boutilier J J, Lee T, Craig T, Sharpe M B and Chan T C Y 2015 Models for predicting objective function weights in prostate cancer IMRT *Medical Physics* **42** 1586-95

Boyd S, Parikh N, Chu E, Peleato B and Eckstein J 2011 Distributed optimization and statistical learning via the alternating direction method of multipliers *Foundations and Trends® in Machine Learning* **3** 1-122

Chan T C Y, Craig T, Lee T and Sharpe M B 2014 Generalized Inverse Multiobjective Optimization with Application to Cancer Therapy *Operations Research* **62** 680-95

Che Z, Purushotham S, Khemani R and Liu Y *AMIA Annual Symposium Proceedings,2016),* vol. Series 2016*)*: American Medical Informatics Association) p 371

Chen L, Shen C, Li S, Maquilan G, Albuquerque K, Folkert M R and Wang J *Medical Imaging: Image Processing,2018),* vol. Series*)* p 1057436

Greenspan H, Van Ginneken B and Summers R M 2016 Guest editorial deep learning in medical imaging: Overview and future promise of an exciting new technique *IEEE Transactions on Medical Imaging* **35** 1153-9

Iqbal Z, Luo D, Henry P, Kazemifar S, Rozario T, Yan Y, Westover K, Lu W, Nguyen D and Long T 2017 Accurate Real Time Localization Tracking in A Clinical Environment using Bluetooth Low Energy and Deep Learning *arXiv preprint arXiv:1711.08149*

Iqbal Z, Nguyen D and Jiang S 2018 Super-Resolution 1H Magnetic Resonance Spectroscopic Imaging utilizing Deep Learning *arXiv preprint arXiv:1802.07909*

LeCun Y, Bengio Y and Hinton G 2015 Deep learning *nature* **521** 436

LeCun Y, Bottou L, Bengio Y and Haffner P 1998 Gradient-based learning applied to document recognition *Proceedings of the IEEE* **86** 2278-324

Lee T, Hammad M, Chan T C Y, Craig T and Sharpe M B 2013 Predicting objective function weights from patient anatomy in prostate IMRT treatment planning *Medical Physics* **40** 121706-n/a Liu H, Shen C, Klages P, Yang M, Albuquerque K, Wu Z, Li J and Jia X 2017 Interactive treatment planning for high dose brachytherapy in gynecological cancer *under preparation*

Lu R, Radke R J, Happersett L, Yang J, Chui C-S, Yorke E and Jackson A 2007 Reduced-order parameter optimization for simplifying prostate IMRT planning *Physics in Medicine & Biology* **52** 849

Ma G, Shen C and Jia X 2018 Low dose CT reconstruction assisted by an image manifold prior *arXiv preprint arXiv:1810.12255*

Mnih V, Kavukcuoglu K, Silver D, Rusu A A, Veness J, Bellemare M G, Graves A, Riedmiller M, Fidjeland A K, Ostrovski G, Petersen S, Beattie C, Sadik A, Antonoglou I, King H, Kumaran D, Wierstra D, Legg S and Hassabis D 2015 Human-level control through deep reinforcement learning *Nature* **518** 529

Moore K L, Brame R S, Low D A and Mutic S *Seminars in radiation oncology,2012),* vol. Series 22): Elsevier) pp 62-9

Nguyen D, Jia X, Sher D, Lin M-H, Iqbal Z, Liu H and Jiang S 2018 Three-Dimensional Radiotherapy Dose Prediction on Head and Neck Cancer Patients with a Hierarchically Densely Connected U-net Deep Learning Architecture *arXiv preprint arXiv:1805.10397*

Nguyen D, Long T, Jia X, Lu W, Gu X, Iqbal Z and Jiang S 2017 Dose Prediction with U-net: A Feasibility Study for Predicting Dose Distributions from Contours using Deep Learning on Prostate IMRT Patients *arXiv preprint arXiv:1709.09233*

Oelfke U and Bortfeld T 2001 Inverse planning for photon and proton beams *Medical dosimetry* **26** 113-24

Shen C, Gonzalez Y, Chen L, Jiang S and Jia X 2018 Intelligent Parameter Tuning in Optimization-Based Iterative CT Reconstruction via Deep Reinforcement Learning *IEEE transactions on medical imaging* **37** 1430

Silver D, Huang A, Maddison C J, Guez A, Sifre L, Van Den Driessche G, Schrittwieser J, Antonoglou I, Panneershelvam V and Lanctot M 2016 Mastering the game of Go with deep neural networks and tree search *nature* **529** 484

Silver D, Schrittwieser J, Simonyan K, Antonoglou I, Huang A, Guez A, Hubert T, Baker L, Lai M and Bolton A 2017 Mastering the game of Go without human knowledge *Nature* **550** 354

Sturm I, Lapuschkin S, Samek W and Müller K-R 2016 Interpretable deep neural networks for single-trial EEG classification *Journal of neuroscience methods* **274** 141-5

Viswanathan A N, Beriwal S, De Los Santos J F, Demanes D J, Gaffney D, Hansen J, Jones E, Kirisits C, Thomadsen B and Erickson B 2012 American Brachytherapy Society consensus guidelines for locally advanced carcinoma of the cervix. Part II: High-dose-rate brachytherapy *Brachytherapy* **11** 47-52

Wahl N, Bangert M, Kamerling C P, Ziegenhein P, Bol G H, Raaymakers B W and Oelfke U 2016 Physically constrained voxel-based penalty adaptation for ultra-fast IMRT planning *Journal of Applied Clinical Medical Physics* **17** 172-89

Wang G 2016 A perspective on deep imaging *arXiv preprint arXiv:1609.04375*

Wang H, Dong P, Liu H and Xing L 2017 Development of an autonomous treatment planning strategy for radiation therapy with effective use of population-based prior data *Medical Physics* **44** 389-96

Watkins C J and Dayan P 1992 Q-learning *Machine learning* **8** 279-92

Webb S 2003 The physical basis of IMRT and inverse planning *British Journal of Radiology* **76** 678-89

Wu X and Zhu Y 2001 An optimization method for importance factors and beam weights based on genetic algorithms for radiotherapy treatment planning *Physics in Medicine & Biology* **46** 1085

Wulfmeier M, Ondruska P and Posner I 2015 Maximum entropy deep inverse reinforcement learning *arXiv preprint arXiv:1507.04888*

Xing L, Li J G, Donaldson S, Le Q T and Boyer A L 1999 Optimization of importance factors in inverse planning *Physics in Medicine and Biology* **44** 2525

Yan H and Yin F-F 2008 Application of distance transformation on parameter optimization of inverse planning in intensity-modulated radiation therapy *Journal of Applied Clinical Medical Physics* **9** 30-45

Yan H, Yin F-F, Guan H-q and Kim J H 2003a AI-guided parameter optimization in inverse treatment planning *Physics in Medicine & Biology* **48** 3565

Yan H, Yin F-F, Guan H and Kim J H 2003b Fuzzy logic guided inverse treatment planning *Medical Physics* **30** 2675-85

Yang Y and Xing L 2004 Inverse treatment planning with adaptively evolving voxel-dependent penalty scheme *Medical Physics* **31** 2839-44

Zhang C, Bengio S, Hardt M, Recht B and Vinyals O 2016 Understanding deep learning requires rethinking generalization *arXiv preprint arXiv:1611.03530*

Zhang Q, Wu Y N and Zhu S-C *The IEEE Conference on Computer Vision and Pattern Recognition (CVPR),2018),* vol. Series*)* pp 8827-36

Zhen X, Chen J, Zhong Z, Hrycushko B, Zhou L, Jiang S, Albuquerque K and Gu X 2017 Deep convolutional neural network with transfer learning for rectum toxicity prediction in cervical cancer radiotherapy: a feasibility study *Physics in Medicine & Biology* **62** 8246

Zhu X, Ge Y, Li T, Thongphiew D, Yin F F and Wu Q J 2011 A planning quality evaluation tool for prostate adaptive IMRT based on machine learning *Med Phys* **38** 719-26